\documentclass[sigconf,authorversion,nonacm]{acmart}

\AtBeginDocument{%
  \providecommand\BibTeX{{%
    \normalfont B\kern-0.5em{\scshape i\kern-0.25em b}\kern-0.8em\TeX}}}

\usepackage[normalem]{ulem}
\usepackage{enumitem}
\usepackage{amsmath,amsfonts}
\usepackage{graphicx}
\usepackage{textcomp}
\usepackage{empheq} 
\usepackage{xcolor}
\usepackage{optidef}
\usepackage[linesnumbered,ruled]{algorithm2e}
\usepackage{multirow}
\usepackage{subfig}
\usepackage{algpseudocode}

\algnewcommand\algorithmicforeach{\textbf{for each}}
\algdef{S}[FOR]{ForEach}[1]{\algorithmicforeach\ #1\ \algorithmicdo}

\def\BibTeX{{\rm B\kern-.05em{\sc i\kern-.025em b}\kern-.08em
    T\kern-.1667em\lower.7ex\hbox{E}\kern-.125emX}}

\begin{document}

\title{Towards Self-Improving Hybrid Elasticity Control of Cloud-Based Software Systems}






\author{Mohan Baruwal Chhetri$^{a,b}$, Abdur Rahim Mohammad Forkan$^b$, Anton V. Uzunov$^{c,a}$, Surya Nepal$^a$}

\affiliation{
	\institution{
	$^a$CSIRO Data61, Australia\\
	$^b$Swinburne University of Technology, Melbourne, Australia\\
	$^c$Defence Science and Technology Group, Adelaide, Australia\\}
}
\email{{Mohan.BaruwalChhetri, Anton Uzunov, Surya.Nepal}@data61.csiro.au, fforkan@swin.edu.au}

\renewcommand{\shortauthors}{Baruwal Chhetri, et al.}

\begin{abstract}
  \textit{Elasticity} is a form of self-adaptivity in cloud-based software systems that is typically restricted to the infrastructure layer and realized through \textit{auto-scaling}. However, both \textit{reactive} and \textit{proactive} forms of infrastructure auto-scaling have limitations, when used separately as well as together. To address these limitations, we propose an approach for self-improving hybrid elasticity control that combines (a) infrastructure and software elasticity, and (b) proactive, reactive and responsive decision-making. At the infrastructure layer, resources are provisioned \textit{proactively} based on one-step-ahead workload forecasts, and \textit{reactively}, based on observed workload variations. At the software layer, features are activated or deactivated in response to transient, minor deviations from the predicted workload. The proposed approach can lead to better performance-aware and cost-effective resource management in cloud-based software systems. 
We validate our approach via a partial realization and simulation with real-world datasets. 
\end{abstract}


\maketitle


\section{Introduction} \label{sec:int}

Self-adaptive software systems are capable of adjusting their run-time behaviour in response to changes stemming from the ``self'' (e.g., failure) and/or the ``environment'' (e.g., increasing workload)~\cite{salehie2009self}. \textbf{Elasticity} is a form of \textit{self-adaptivity} -- or more precisely, a combination of self-configuration and self-optimization (cf. \cite{alDhuraibi2017elasticity}) -- in cloud-based systems that is typically restricted to the infrastructure layer. It is defined as a system's ability to ``\textit{adapt to workload changes by provisioning and deprovisioning resources in an autonomic manner, such that at each point in time the available resources match the current demand as closely as possible}''~\cite{herbst2013elasticity}.
The aforementioned process of automatically provisioning or deprovisioning computing resources is termed \textbf{auto-scaling} (see \cite{chen2018survey, alDhuraibi2017elasticity}). 

There are two main approaches for auto-scaling of cloud infrastructure: \textit{reactive} and \textit{proactive}. Using \textbf{reactive scaling}, 
reconfiguration actions are triggered after the system has been subjected to stress, potentially resulting in some performance degradation and SLA violations. The subsequent provisioning of resources takes time resulting in further delays in converging to the required resource levels. Using \textbf{proactive scaling}, the provisioning of resources is scheduled in advance based on the forecast workload and the corresponding resource requirements \cite{chhetri2019exploiting}. Hence, it addresses the main limitations of reactive scaling. 
However, forecasting is not an exact science, and therefore, situations may arise, even with conservative resource allocations, where the resources provisioned to handle the \textit{predicted} workload are insufficient to handle the \textit{actual} workload. 
The shortcomings of using reactive or proactive scaling in isolation can be overcome by combining both so that resources are proactively scheduled based on the forecast workload 
and any additional resource adjustments are made reactively. However, the issue of \textbf{resource thrashing}, which is a ``temporary, yet very quick, oscillation in the allocation of resources''~\cite{bersani2014towards}, remains. Thrashing is caused by oscillating scaling decisions, which can negatively impact the system performance and lead to increased (wasted) cloud spend. 

To address these shortcomings, we propose a self-improving\footnote{Self-improvement is a special \mbox{self-*} property that implies ``\textit{adjustment of the adaptation logic}''~\cite{krupitzer2016comparison}, thereby subsuming self-adaptivity.} hybrid elasticity controller for cloud-based software systems that combines (a) infrastructure elasticity with software elasticity, and (b) proactive, reactive and responsive decision-making via a hierarchical control-loop architecture. 
\textbf{Software elasticity} is the capability of a cloud-based software system to adapt itself at the software layer in an autonomic manner by activating and deactivating features in response to transient workload fluctuations~\cite{dupont2015experimental}. Compared to infrastructure elasticity, software elasticity is much more responsive and a lot cheaper to effect. 
Using our controller, 
resources are proactively provisioned at the infrastructure layer based on one-step-ahead workload forecasts. Transient, minor deviations from the forecast workload are handled responsively\footnote{Responsive decision-making combines proactive and reactive decision-making over smaller time-scales.} by activating and deactivating application features at the software layer to maintain service reliability and performance. Larger, sustained deviations are handled through reactive scaling at the  infrastructure layer. To the best of our knowledge, ours is the first approach to jointly leverage infrastructure and software elasticity and combine proactive, reactive and responsive decision-making. 

An initial experimental validation of our approach, using a partial realization of the hybrid controller (inclusive of concrete auto-scaling algorithms), and simulation with real-world datasets, confirms that a hybrid controller can exhibit significant improvements in request handling performance. 

\begin{figure*}[!t]
	\centering
	\includegraphics[width=0.85\hsize]{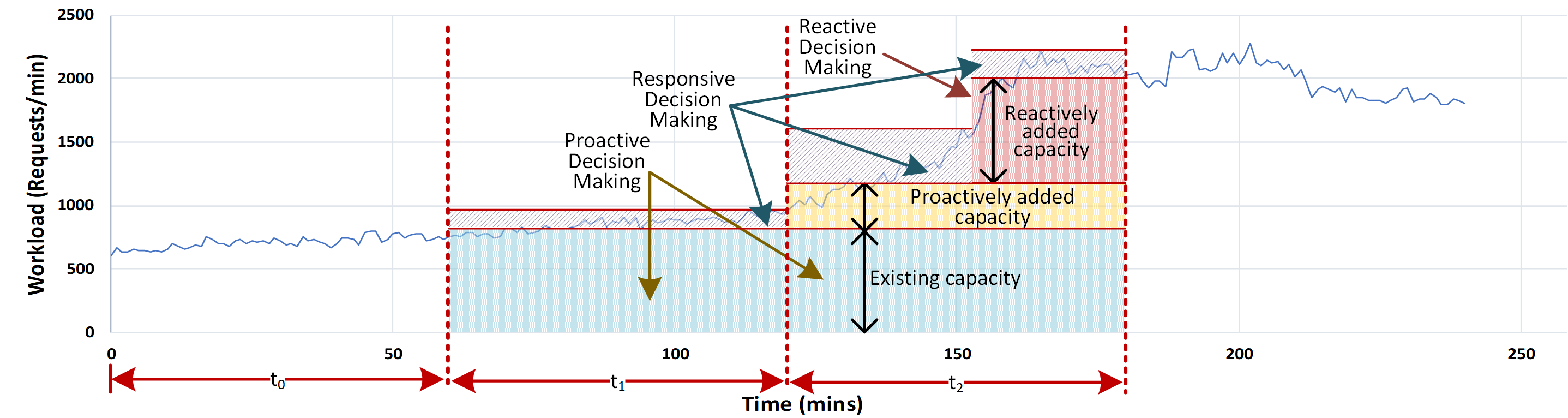} 
	\caption{Proactive, Reactive and Responsive Elasticity Control.}
	\label{fig:proreares}
\end{figure*}

The rest of the paper is organized as follows. Section~\ref{sec:relwork} briefly discusses related work on self-adaptive elasticity control in cloud-based software systems. Section~\ref{sec:sysdes} presents an overview of our proposed self-improving hybrid elasticity controller. We present a realization of the controller in Section~\ref{sec:hecr}, and evaluate its request handling performance in Section~\ref{sec:eval}. 
Section~\ref{sec:conc} concludes the paper with a discussion of future work.


\section{Related Work} \label{sec:relwork}

According to Rightscale~\cite{rightscale2020}, 35\% or more of cloud spend is ``\textit{wasted spend}'' due to resource allocation inefficiencies. Yet, very few organizations have implemented automated processes for resource management to address these issues and optimizing cloud costs continues to be a top initiative for cloud users of all maturity levels~\cite{rightscale2020}. 
Most research on elasticity management in cloud-based systems has focused on combining reactive and proactive elasticity to overcome the limitations of using either approach in isolation~\cite{ali2012adaptive, liu2015prorenata, hirashima2016proactive, moore2013coordinated}. However, just combining proactive and reactive infrastructure scaling does not address the issue of resource thrashing.

Klein et al.~\cite{klein2014brownout} first proposed the self-adaptation programming paradigm of \textbf{brownout} as a complementary approach to infrastructure elasticity so that cloud applications can withstand unpredictable runtime variations without over-provisioning. Brownout is based on the notion that \textit{optional code can be dynamically activated/deactivated in response to resource constraints so that throughput and response times are maintained at the expense of user experience}. More recently, this concept has been extended further by introducing the notion of \textit{quality-elasticity}~\cite{larsson2019quality} so that self-adaptation at the software layer is not restricted to a binary decision (feature on/off) but can occur on a more fine-grained spectrum. 

In~\cite{dupont2015experimental}, the authors present an experimental analysis on autonomic strategies for cloud elasticity. This is one of the first papers that uses the notion of cross-layer elasticity for cloud-based applications and advocates leveraging software elasticity to overcome the shortcomings of infrastructure elasticity. It distinguishes between horizontal and vertical software scaling and also presents experimental results combining infrastructure and software elasticity. However, it only considers reactive scaling of infrastructure resources. 

In~\cite{dustdar2011principles}, the authors define three types of elasticity -- \textit{resource elasticity}, \textit{cost elasticity}, and \textit{quality elasticity}. While resource elasticity refers to adjustments to computing resources, cost elasticity describes a resource's responsiveness to changes in cost and quality elasticity measures how responsive quality is to a change in resource usage.  
In~\cite{chhetri2019exploiting}, the authors propose an approach for proactive resource scaling that considers resource and cost elasticity; cost-optimal resources are scheduled in advance based on the one-step-ahead forecast workload and forecast resource prices and any shortcomings are handled through reactive scaling. However, the issue of transient workload fluctuations is not addressed adequately. 

Multiple control loops for auto-scaling, including multiple MAPE-K loops, have been used by various approaches in the literature (see \cite{chen2018survey}); however, none of these propose our combination of proactive/reactive/responsive decision-making across both the infrastructure and software levels. 
Our proposed approach complements our previous work in~\cite{chhetri2019exploiting} and makes it more cost-effective and performance-efficient. As a simple illustrative example, rather than allocating resources conservatively at the infrastructure layer, i.e., for the maximum forecast workload, the proactive elasticity controller can allocate them at a lower percentile and handle any workload deviations through software scaling leading to increased cost savings. 

\section{Hybrid Elasticity Controller Overview} \label{sec:sysdes}

In this section, we present an overview 
of our proposed 
hybrid elasticity controller. 
First, we provide a simple motivating scenario that serves as a running example to explain infrastructure and software scaling as well as proactive, reactive and responsive decision-making. Next, we provide an overview of the controller's conceptual architecture, 
followed by brief descriptions of how the three types of decision-making 
work together to realize elasticity. 

\subsection{Motivating Scenario} \label{sec:ms}

Consider an e-commerce website (a more-or-less ``canonical'' example from the literature) that, in addition to the standard search, select and purchase functionality, also offers end-users recommendations of similar products they might be interested in. We assume that \mbox{Fig.~\ref{fig:proreares}} represents the requests received by such a  website (the workload representation in the figure is realistic, generated from the Wikipedia 2007 (Wiki07\footnote{\url{http://www.wikibench.eu/?page_id=60}}) access trace logs. The recommender engine greatly enhances the user experience, but is resource-intensive. Ideally, the  website should serve as much content as possible to maximize quality of user experience, while maintaining high levels of quality of service (QoS) -- prioritizing throughput and response times in particular. Achieving this goal clearly requires elasticity. Specifically, an auto-scaling controller would need to be implemented such that both long-lived/substantial, and transient/small deviations in workloads are handled appropriately over different time-periods without compromising the website's functional and non-functional requirements.

\begin{figure}[!t]
	\centering
	\includegraphics[width=1.0\hsize]{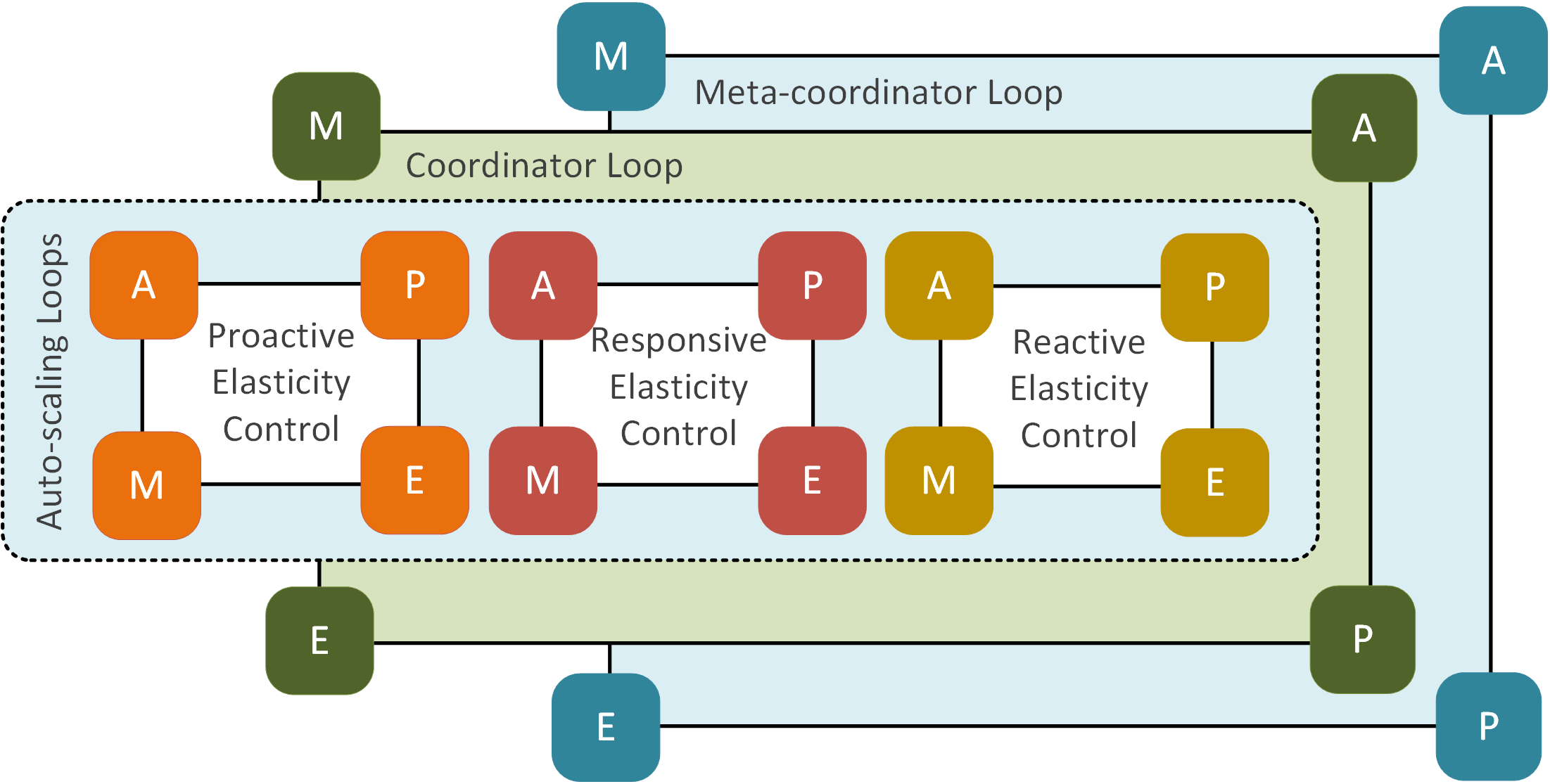} 
    \caption{Hierarchical Control Loop Architecture of Hybrid Elasticity Controller.}
	\label{fig:pis}
\end{figure}

\subsection{Controller Architecture} \label{sec:ca}

Conceptually, any self-adaptive system comprises two constituent parts: the \textit{application logic}, realizing the domain functionality, and the \textit{adaptation logic} that is responsible for the runtime adaptation of the application logic. In the context of cloud-based software systems, an elasticity controller can be considered as the adaptation logic that operates in an externalised form, implementing the popular \mbox{\textit{MAPE-K}} reference model~\cite{computing2006architectural} to realize \textit{self-configuration} and \textit{self-optimization}. In our hybrid controller, this single, ``conceptual'' MAPE-K loop is implemented by a three-fold hierarchy of (more concrete) MAPE-K loops, as shown in Fig.~\ref{fig:pis}: three \textit{auto-scaling loops}, each one using proactive, responsive and reactive decision-making, respectively; one \textit{coordinator loop} to control and adapt the three auto-scaling loops; and one \textit{meta-coordinator loop} to adapt the coordinator loop itself. The purpose of the coordinator loop is to ensure consistency among the decisions made by the auto-scaling loops -- thereby avoiding conflicts -- and also to help realize self-improvement. The purpose and operation of the auto-scaling loops is outlined in the next three sub-sections. 

\subsubsection{Proactive Infrastructure Control (PIC)}

The first control loop in our hybrid elasticity controller is associated with proactive infrastructure scaling. 
With reference to Fig.~\ref{fig:proreares}, \textbf{proactive decision-making} is used in time period t\textsubscript{n} to determine the infrastructure requirements in time period t\textsubscript{n+1} for responding fully to requests, i.e., with both the mandatory and optional parts, based on the forecast workload. For instance, the resource requirement for time period $t\textsubscript{2}$ is proactively determined in time period $t\textsubscript{1}$. The typical duration of the one-step ahead time period is 1 hour because most IaaS providers use the per instance-hour PAYG pricing model. 

The MAPE-K elements work as follows. The \textit{Workload Monitor} monitors the incoming requests in real-time (\textbf{M}). 
The \textit{Workload Estimator} computes an estimation of the one-step ahead workload (\textbf{A}) based on the historical workload data and any other relevant knowledge (\textbf{K}) that might impact the forecast. The Workload Estimator uses an appropriate look-back horizon (e.g., 24 hours, 7 days, 30 days etc.) to forecast the workload for the next hour. 
The forecast workload is forwarded to the \textit{Infrastructure Optimizer}, which computes the corresponding ``cost-optimal'' resource allocation (\textbf{P}) based on the current resource allocation (obtained through the \textit{Infrastructure Monitor}), the sustainable workload capacity of the candidate instance types\footnote{Sustainable workload capacity is the maximum workload that an instance type can sustain over time without violating the QoS.} (typically obtained through performance benchmarking), and the real-time pricing information (\textbf{K}) for the candidate instance types (obtained through the \textit{Price Monitor}). Finally, the output of the Infrastructure Optimizer, i.e., the \textit{reconfiguration plan}, is forwarded to the \textit{Infrastructure Orchestrator}, which is responsible for effecting the reconfiguration of the underlying infrastructure 
(\textbf{E}). 

The main benefit of proactive decision-making is that it gives the elasticity controller sufficient time to prepare resources in anticipation of future workload changes. However, as discussed in Section~\ref{sec:int}, workload forecasting is not an exact science and the effectiveness of PIC depends upon the accuracy of the forecasting algorithm/s. 

\subsubsection{Responsive Software Control (RSC)}

The second control loop in our hybrid elasticity controller is associated with responsive software scaling, realizing the brownout approach of \cite{klein2014brownout}. As can be seen in Fig.~\ref{fig:proreares}, there are several time points in 
periods t\textsubscript{1} and t\textsubscript{2}  where the actual workload is higher than that which can be handled by the proactively allocated resources. \textbf{Responsive decision-making} at the software layer can deal with these transient, minor deviations from the predicted workload via feature variation. 
%
For example, the \textit{Workload Estimator}, (\textbf{A})nalyse element of the RSC loop, may predict the next 1 minutes' workload based on the workload observed by the \textit{Workload Monitor} in the past 5 minutes' (\textbf{M}). If the ``newly'' estimated transient workload cannot be handled by the available resources, then a software reconfiguration action is triggered by the \textit{Software Optimizer} (\textbf{P}) so that ``optional'' features are deactivated by the \textit{Software Configurator} (\textbf{E}). Similarly, if the estimated workload is less than the predicted workload, additional/optional features that have been deactivated can be reactivated. Returning to our motivating scenario, an example reconfiguration action is to disable a web-site's recommender system, thereby reducing the richness of information presented to users (quality of experience) but maintaining the same QoS levels (throughput and response time). 
As proposed in~\cite{larsson2019quality}, the quality of the response can be modified along a \textit{fine-grained spectrum} subject to the availability of resources for handling the transient workload deviations. 

Responsive decision-making allows transient, minor workload deviations to be absorbed as they happen, without 
triggering expensive, 
potentially conflicting, infrastructure reconfiguration actions, thereby avoiding resource thrashing.

\subsubsection{Reactive Infrastructure Control (RIC)}

The third control loop in our proposed elasticity controller is associated with reactive infrastructure scaling. 
\textbf{Reactive decision-making} occurs when the deviations in the observed workload are much larger and sustained, and cannot be addressed through software elasticity control. 
With reference to Fig.~\ref{fig:proreares}, in time period $t\textsubscript{2}$, the workload deviations are larger and more sustained, triggering the associated scaling rule as a result of which additional resources (as indicated by the red box) are reactively added to the existing infrastructure capacity (indicated by the blue and yellow boxes). These scaling rules are defined around the ``observed'' workload as opposed to other system metrics such as CPU or memory of the provisioned instances. 
Additionally, reactive infrastructure scaling can also be triggered by price changes, e.g., if using Amazon EC2, Spot instance termination due to Spot price changes can trigger reactive scaling at the infrastructure layer. These particular scaling rules are typically defined by domain experts but can also be learnt. 
The MAPE-K elements for RIC are similar to those for PIC. 

Reactive scaling can handle the inaccuracies in the proactive resource scaling by making reactive adjustments to the allocated resources. However, it should be noted that there is still a need for responsive elasticity control at the software layer to handle the minor workload deviations,i.e., reactive and proactive infrastructure scaling and responsive software scaling are complementary. 


\subsection{Realizing Self-Improvement}

As can be inferred from the preceding exposition, self-configuration and self-optimization are realizable at the infrastructure layer via the PIC and RSC loops, and at the software layer by the responsive ``feature management'' control loop. \textit{Self-improvement}, which refers to the improvement of the self-* capabilities through adjustment of the adaptation logic, can be realized across all three auto-scaling loops via the coordinator loop -- and within the coordinator loop itself via the meta-coordinator control loop -- as follows: 

\begin{itemize}[leftmargin=*]
    \item \textit{Proactive auto-scaling}: meta-learning could be used to facilitate improved prediction, e.g., by fine-tuning the parameters of a specific prediction algorithm, or supporting the selection of the best algorithm from among several candidates for better decision-making~\cite{krupitzer2018satisfy}.
    
    \item \textit{Reactive auto-scaling}: the threshold values for existing scaling rules could be fine-tuned, and new scaling rules could also be learnt online. 
    
    \item \textit{Responsive auto-scaling}: if software variability is captured initially as a feature model where features can be turned on and off, then it should be possible to generate the (application-specific) adaptivity/reconfiguration code on-the-fly. Further to this, the responsive auto-scaling loop could be configured (by the coordinator loop) to select among multiple reconfiguration strategies/plans beyond simple feature on/off, learning which one gives best performance/QoS.
    
    \item \textit{Coordination control}: as for proactive auto-scaling, the algorithms used for optimization can be changed at runtime, and the (meta-level) rules which generate the scaling rules for reactive, responsive and proactive decision-making can also be modified. 
\end{itemize}

\section{Hybrid Elasticity Controller Realization} \label{sec:hecr}

\begin{figure}[!t]
\centering
\includegraphics[width=1.0\hsize]{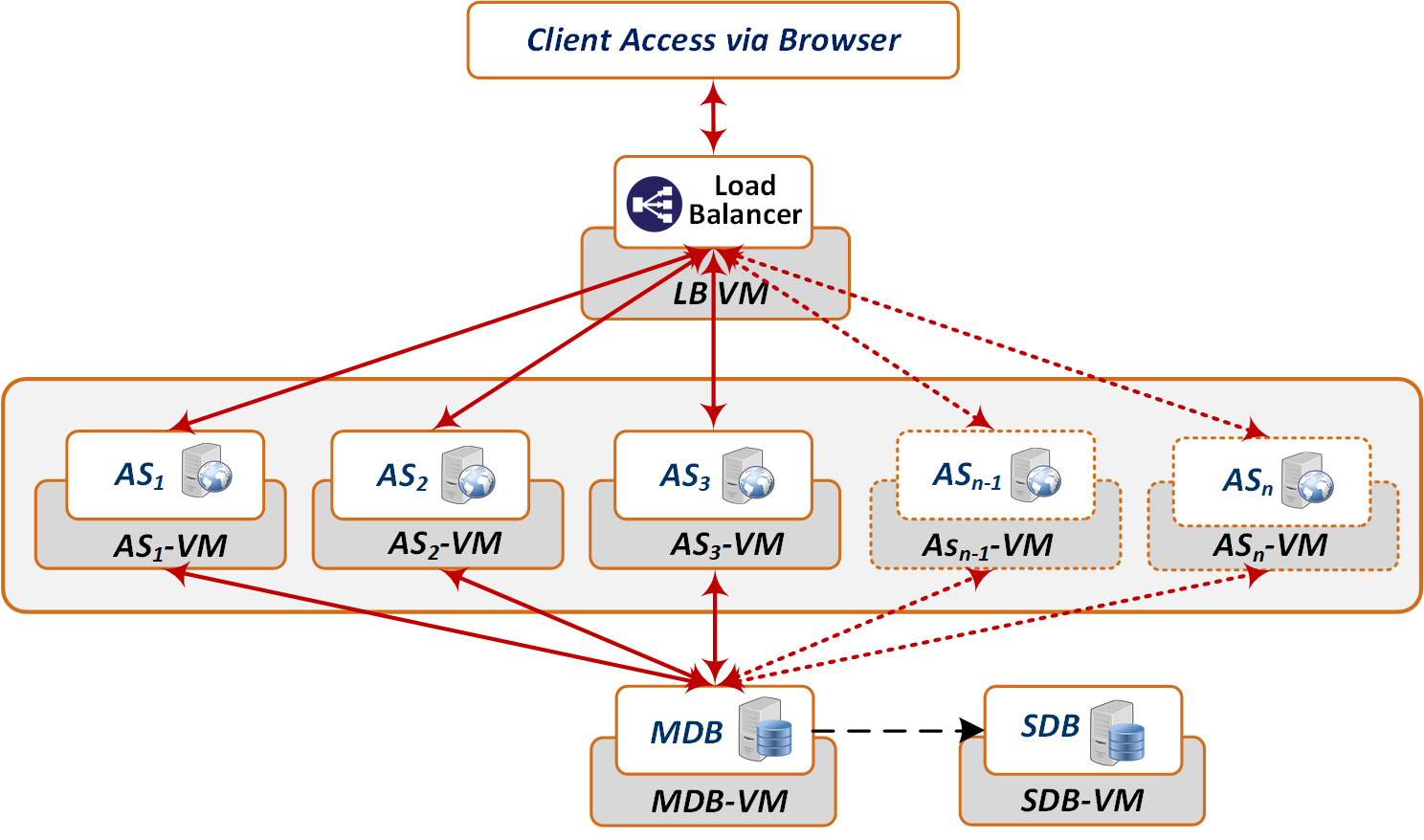}
\caption{Simplified multi-tier Cloud-based Software System.} \label{fig:arch}
\end{figure} 

In this section, we present a concrete, partial realization of the hybrid elasticity controller architecture proposed in Section~\ref{sec:sysdes} -- realizing, in particular, all the auto-scaling inner loops and a (nominally) hard-coded/manual version of the coordinator loop, but omitting the meta-coordinator loop. To this end, we first present a workload estimation model that is used in both the PIC and RSC loops. We then describe how the PIC, RSC, and RIC loops are realized, using a typical cloud-based transactional web application as the managed application (see Fig.~\ref{fig:arch}) to illustrate how resources at the logic tier are managed through proactive, reactive and responsive auto-scaling. 

\subsection{Workload Estimation Model} \label{sec:em}

The \textbf{A}nalyse step of both the PIC and RSC loops uses one-step-ahead workload estimation based on time series forecasting. In our workload estimation model, the historical workload at time point $t$ for a lookback horizon $l$ is denoted by $\mathcal{W}_{t-l} = \{w_{t-(l-1)}, \dots, w_{t-1}, w_t\}$. Similarly, the forecast workload for forecast horizon $f$ is denoted by $\hat{\mathcal{W}}_{t+f} = \hat{w}_{t+1}, \hat{w}_{t+2}, \dots, \hat{w}_{t+f}$. If the forecasts are derived for duration $f$ using look back $l$ and forecasting model $k$, then the forecast workload is denoted by $\hat{\mathcal{W}}_{t+f|t-l,k}$. 

There are two key differences in how workload estimation works in PIC and RSC.
\begin{itemize}[leftmargin=*]
    \item The first difference is in the time scales used in the two control loops. For PIC, the forecast duration is in hours (e.g., 1 hr, 2 hrs) with a lag of 1 minute and the lookback duration is one or more hours. For RSC, the forecast duration is typically a few seconds or minutes with a lag of 1 second, and the lookback duration is less than the PIC forecast duration (e.g. 1 hr). 
    \item The second difference is in the output returned by the workload estimation model. In the PIC loop, the workload estimation model returns a single value that is used for computing the optimal hourly resource allocation. However, in the RSC loop, the model returns all the predicted value/s, which are used by the \textit{Software Reconfigurator}. Thus, in the RSC loop the model returns $\hat{\mathcal{W}}_{t+f|t-l,k}$, whereas in the PIC loop it returns the one-step-ahead workload estimation $\hat\omega_{t+}$, given by:
    \begin{equation} \label{eqn:wef}
        \hat{\omega}\textsubscript{t+} = \phi(\hat{\mathcal{W}}_{t+f|t-l,k})
    \end{equation}
where $\phi \in$ \text{\{max, mean, median, n\textsuperscript{th} percentile \dots \}} determines the one-step-ahead workload value. Conservative estimation will pick the maximum value from the predicted workload whereas a less conservative approach may choose an appropriate value less than the maximum, e.g. 95\textsuperscript{th} percentile. 

\end{itemize}

\subsection{Proactive Infrastructure Control} \label{sec:pras}


As mentioned in Section~\ref{sec:sysdes}, the PIC loop is responsible for proactive resource allocation based on the estimated one-step-ahead workload. In our realization of the PIC loop, we model cloud infrastructure resource allocation as an optimisation problem and, for a given workload estimate, use the \textit{delta capacity optimization} (DCO) strategy, which we have previously proposed in \cite{chhetri2019exploiting}, to perform proactive resource scaling. The main objective of this strategy is to scale resources based on the delta (i.e., difference) between the current allocated capacity in the time period $t$ and the required capacity in the next time period $t+$. 

The DCO strategy is based on two assumptions. The first assumption is that the \textit{sustainable workload capacity} for each VM type is known \textit{a priori}, either through benchmarking or other means~\cite{chhetri2019exploiting}. The second assumption is that the aggregate sustainable capacity of all VMs at the logic tier is equal to the sum of the individual workload capacities. Based on these two assumptions, infrastructure resources are proactively scaled up or down as described below.

\subsubsection{Delta Capacity Optimisation for Scale-out.} 

Equation~\ref{eqn:so} computes the cost-optimal resource capacity that must be added to the existing capacity if the workload is predicted to increase in the next time period. 
\begin{equation} \label{eqn:so}
    \begin{split}
    \text{\textit{minimize}} \sum_{i \in \mathcal{I}}{({\gamma_{i}}{x_i})}_{t+} \\
    \text{s.t} \; \sum_{i \in \mathcal{I}}{({\lambda_{i}}{x_i})}_{t+} \gtrapprox \delta_{t \mid t+} \\
    \text{where} \; \delta_{t \mid t+} = \hat{\omega}_{t+} - \mathcal{R}_{t} 
    \end{split}
\end{equation} 
    
\begin{itemize}
    \item $\mathcal{I} = \{1, 2, \dots, n\}$ denotes the set of candidate instance types,
    \item $\gamma_i$ denotes the cost per unit time of $i \in \mathcal{I}$,
    \item $\lambda_i$ denotes the sustainable workload capacity of $i \in \mathcal{I}$,
    \item $x_i \geq 0$ denotes the number of instances of $i \in \mathcal{I}$,
    \item $\mathcal{R}_{t} = \sum\limits_{i \in I}{(\lambda_{i}{x_{i}})_{t}}$ denotes the allocated capacity in time period $t$, 
    \item $t+$ denotes the next time step,
    \item $\hat{\omega}_{t+}$ denotes the one-step-ahead predicted workload, and
    \item $\delta_{t\mid t+}$ denotes the delta between the one-step-ahead predicted workload and the existing resource capacity.
\end{itemize}  

\subsubsection{Delta Capacity Optimisation for Scale-in.} Equation~\ref{eqn:si} computes the optimal resource capacity that must be removed from the existing capacity if the workload is predicted to decrease in the next time period. 
\begin{equation} \label{eqn:si}
    \begin{split}
    \text{\textit{maximize}}\sum_{j \in \mathcal{J}}{({\gamma_{j}}{x_j})}_{t+} \\
    \text{s.t} \; \sum_{j\in \mathcal{J}}{({\lambda_{j}}{x_j})}_{t+} \lessapprox \delta_{t|t+} \\
    \text{where} \; \delta_{t|t+} = \mathcal{R}_t - \hat{\omega}_{t+} \\
    \text{and} \; \mathcal{J} \subseteq \mathcal{I} 
    \end{split}
\end{equation} 
$\mathcal{J}$ denotes the set of instance types included in the current capacity.
\subsubsection{PIC Algorithm}

Algorithm~\ref{alg:pic} summarises the two-step process for PIC. First, the one-step-ahead workload is estimated based on the historical workload information (line 1). Next, the delta between the estimated workload and the aggregate sustainable capacity of currently allocated resources is computed (line 2). If additional resources are required, Equation~\ref{eqn:so} is used to compute the optimal allocation and their startup is scheduled so that the resources are ready for use in the next time period (lines 3-5). Similarly, if resources have to be removed, Equation~\ref{eqn:si} is used to compute the optimal deallocation from the existing capacity (lines 6-8). If there is no change in the workload, then no action is taken and the system persists with the existing resource allocation (lines 9-10).  

\begin{algorithm}[t]
	\footnotesize
	\SetKwInOut{Input}{Input}
	\SetKwInOut{Output}{Output}
	
	\Input{Historical workload $\mathcal{W}_{t-l}$, forecasting model $k$ and existing resource allocation $\mathcal{R}_{t}$ for the PIC loop} 
	\Output{Optimal one step ahead resource allocation}
	\BlankLine
	Compute $\hat{\omega}_{t+}$ using Equation~\ref{eqn:wef} \\
	Compute $\delta_{t|t+} = \hat{\omega}_{t+} - \mathcal{R}_{t}$ \\

	\uIf{$\delta_{t|t+} > 0$}{
         Determine resources to add to $\mathcal{R}_{t}$ using Equation~\ref{eqn:so} \\
         Schedule their spin up in time period $t+$ \\
    }
    \uElseIf{$\delta_{t|t+} < 0$}{
        Determine resources to remove from $\mathcal{R}_{t}$ using Equation~\ref{eqn:si} \\
        Schedule their shut down in time period $t+$ 
    }
	\Else{
        Persist with $\mathcal{R}_{t}$ 
    }
    \caption{PIC Algorithm} \label{alg:pic}
\end{algorithm} 

\subsection{Responsive Software Control} \label{sec:ras}

In realizing RSC, we assume that the application shown in Fig.~\ref{fig:arch} is a reconfigurable or brownout-compliant application as proposed in~\cite{klein2014brownout} so that each request can be handled in two ways: (i) \textit{fully}, in which case the response contains both the required and optional content, or (ii) \textit{partially}, in which case only the required content is returned. We also assume that it is possible to activate/deactivate the optional computation on demand, and the associated cost is negligible. 

We extend the idea of sustainable capacity for reconfigurable/ brownout-compliant applications so that in time step $T$ of the PIC loop, the resources at the logic tier $\mathcal{R}_T$ can either \textit{respond fully} to $\alpha_T$ requests or \textit{respond partially} to $\beta_T$ requests, \textit{i.e.,} \textit{$\alpha_T$ and $\beta_T$ form the lower and upper bounds, respectively, on the sustainable throughput in time step $T$}. For simplification, we make the assumption that irrespective of the size and composition of $\mathcal{R}_T$, 
the ratio $\frac{\beta}{\alpha}$ remains constant, \textit{i.e.,} one fully handled request is equivalent to $\frac{\beta}{\alpha}$ partially handled requests.

\subsubsection{Software Scaling Problem} 

Within the RSC loop, the objective is to adapt to the transient workload variations at the software layer so that maximum number of requests can be processed while maintaining the required response time. In time step $t+$, let $\zeta_{t+}$ denote the number of requests that can be processed fully, $\eta_{t+}$ denote the number of requests that can be processed partially, and $\theta_{t+}$ denote the number of requests that cannot be handled due to resource constraints. The key decision in the \textbf{A}nalyse phase of the RSC loop is to then determine the optimal one-step-ahead \textbf{$\zeta_{t+}\mid\eta_{t+}\mid\theta_{t+}$} split in each time step $t$. Three scenarios are possible depending upon the estimated workload $\hat{\omega}_{t+}$:


\begin{itemize}[leftmargin=*]
    \item \textit{Scenario 1 ($\hat{\omega}_{t+} \leq \alpha_T$).} If this condition holds true, then, assuming that the forecast is accurate, the existing resource capacity $\mathcal{R}_T$ can fully respond to all requests, \textit{i.e.,} $\zeta_{t+} = \hat{\omega}_{t+}$. However, forecasting is not an exact science. Therefore, we set $\zeta_{t+} = \alpha_T$, \textit{i.e.,} since the resource capacity can handle up to $\alpha_T$ requests fully. 
    
    \item \textit{Scenario 2 ($\hat{\omega}_{t+} > \beta_T$).} If this condition holds true, then the best possible adaptive action is to \textit{respond partially to $\beta_T$ requests}, \textit{i.e.,} maximize the number of handled requests, so that $\zeta_{t+} = 0$, $\eta_{t+} = \beta_T$ and $\theta_{t+} = \hat{\omega}_{t+} - \beta_T$. The system can still handle $\beta_T$ requests even if the real workload exceeds the predicted one. However, if the real workload $\omega_{t+}$ is less than $\beta_T$ (or $\alpha_T$), then the optimal outcome is not achieved since some requests that could have been fully handled, are instead handled partially. 
    
    \item \textit{Scenario 3 ($\alpha_{T} < \hat{\omega}_{t+} \leq \beta_T$).} If this condition holds true, then the system can handle all requests, some fully and the remaining partially. The adaptive action is to \textit{maximise the number of fully handled requests $\zeta_{t+}$ while ensuring that all requests are handled}. 
    \begin{equation}\label{eqn:rsc}
    \begin{split}
        \text{\textit{maximize}} \; \zeta_{t+} \\ 
        \text{s.t} \; \zeta_{t+} + \eta_{t+} = \hat{\omega}_{t+} 
    \end{split}
    \end{equation} 
    
    As with the previous two scenarios, the effectiveness of the scaling action depends upon the accuracy of the forecast. If $\omega_{t+}$ is less than $\alpha_T$, then there is no negative impact on the system performance. However, if $\omega_{t+}$ is greater than $\hat{\omega}_{t+}$, some requests will not be handled by the system. 
\end{itemize} 

Algorithm~\ref{alg:rsc} summarizes the key steps involved in determining the  $\zeta_{t+}\mid\eta_{t+}\mid\theta_{t+}$ split under the three scenarios outlined above. In particular, it provides a specific approach for determining the optimal workload split between $\zeta_{t+}$ and $\eta_{t+}$ (lines 7-11).   

\begin{algorithm}[t]
	\footnotesize
	\SetKwInOut{Input}{Input}
	\SetKwInOut{Output}{Output}
	\SetKw{KwBy}{by}
    \SetKwFor{For}{for (}{) $\lbrace$}{$\rbrace$}
	
	\Input{Lower bound $\alpha_T$, Upper bound $\beta_T$, Estimated Workload $\hat{\omega}_{t+}$}
 	\Output{Workload split $\zeta_{t+}\mid\eta_{t+}\mid\theta_{t+}$}
 	\BlankLine
 	
 	$\zeta_{t+} \gets 0$;  \\
    $\eta_{t+} \gets 0$; \\
    $\theta_{t+} \gets 0$; \\
 	
   \uIf{$(\hat{\omega}_{t+} \leq \alpha_T)$} {
        $\zeta_{t+} \gets \alpha_T$;
    }
    \uElseIf{$(\hat{\omega}_{t+} > \beta_T)$} {
        $\eta_{t+} \gets \hat{\beta}_{T}$; 
        $\theta_{t+} \gets \hat{\omega}_{t+} - \beta_T$; 
    }
    \uElseIf{$(\alpha_{T} < \hat{\omega}_{t+} \leq \beta_T)$}{
       \For{$i \gets \alpha_T;\ i \geq 0;\ i \gets i-1$} {
           \uIf{($i+\frac{\beta}{\alpha}(\alpha_T - i) \geq \hat{\omega}_{t+}$)} {
               $\zeta_{t+} \gets i$; \\
               $\eta_{t+} \gets \hat{\omega}_{t+} - i$; \\
               \textbf{break};
           }
        }
	}

    \caption{RSC Algorithm} \label{alg:rsc}
\end{algorithm}


\subsection{Reactive Infrastructure Control} \label{sec:ris}

The RIC loop handles workload deviations that are much larger and sustained, and therefore cannot be addressed in the RSC loop. In the \textbf{A}nalysis phase of the RIC loop, the \textit{Policy Engine} evaluates the actual workload, as observed by the \textit{Workload Monitor} (\textbf{M}), against predefined scaling policies. If the observed workload exceeds or drops below certain thresholds for sustained periods, then the corresponding scaling action is triggered. The scale-out policy is triggered when ``the monitored workload $\omega_t$ exceeds a threshold $\varphi_{so}$, $n$ number of times within an evaluation period $t_{e}$''. Similarly, the scale-in policy is triggered when ``the monitored workload drops below the threshold $\varphi_{si}$, $n$ number of times within an evaluation period $t_{e}$.'' $\varphi_{so}$, $\varphi_{si}$, $n$ and $t_{e}$ are typically predefined and static. 

In our specific realization of the RIC loop, the upper and lower bounds on the sustainable capacity for all candidate instance types are known a priori as explained in Section~\ref{sec:ras}, \textit{i.e.,} $\forall{i} \in \mathcal{I}, \lambda_i$ and therefore $\alpha_i$ and $\beta_i$ are known. 
Assuming that at time step $T-1$, the PIC loop computes the resource allocation $\mathcal{R}_T$ for time step $T$, we set the threshold for scale-out $\varphi_{so}$ and scale-in $\varphi_{si}$ as follows: $\theta_{so} = \mathcal{R}_T + \kappa \min\limits_{i \in \mathcal{I}}\lambda_i$ and $\theta_{si} = \mathcal{R}_T - \kappa \min\limits_{i \in \mathcal{I}}\lambda_i$ where $\min\limits_{i \in \mathcal{I}}\lambda_i$ refers to the smallest sustainable capacity in $\mathcal{I}$ and $\kappa$ determines the threshold value. Thus, we define one scale-out policy and one scale-in policy, which get triggered if the monitored workload exceeds $\varphi_{so}$ or drops below $\varphi_{si}$ $n$ number of times within the evaluation period $t_e$. 
However, in contrast to existing approaches, the scale-in/out policies do not explicitly specify which instance type to remove and how many of them. Instead, depending upon the observed change in the workload $\delta_{t \mid t+}$ = $|\mathcal{R}_T - \omega_{t+}|$, the same approach used in the PIC loop (cf.~Section~\ref{sec:ras}) is used to determine the optimal resource capacity that should be reactively added to or removed from the existing capacity.  

\subsection{Control Loop Coordination} \label{sec:controller}

As per our hybrid controller architecture, a coordinator loop is required to set the initial parameters of the individual auto-scaling loops and to reconfigure them as necessary for self-improvement, resolving conflicts along the way. 
In our current controller realization we conceptualize the coordinator loop as a simple rule-set which generates an initial configuration that by default avoids any potential conflicts: \textit{i.e.,} we assume pre-set parameters such as the look-back and look-ahead time windows in the PIC and RSC realizations, the models selected for workload forecasting, and the function to estimate the one-step-ahead workload based on the forecasts. This also implies that the coordinator loop is static and hence all reconfiguration changes must be brought about manually, which is sufficient for the scope of our validation. 


\section{Experimental Validation} \label{sec:eval}

In this section, we validate the hybrid elasticity controller realization presented in Section~\ref{sec:hecr}. We first list the key assumptions that we have made in our experiments and then validate the PIC, RSC and RIC loops via simulations. 

\begin{table}[!t]
  \centering
   \caption{Candidate Instance Types.} 
    \begin{tabular}{c|c|c} 
    \hline
      \textbf{Instance} & \textbf{Sustainable} &  \textbf{On-Demand Price }\\
      \textbf{Type} & \textbf{Capacity} ($\alpha$) & \textbf{(AUD)} \\
      \hline
      t2.medium & 200 & 0.0584\\
      t2.large & 400 & 0.1168\\
      t2.xlarge & 700 & 0.2336\\
      m4.large & 400 & 0.125 \\
      m4.xlarge & 750 & 0.25 \\
      c4.large & 500 & 0.13 \\
      c4.xlarge & 750 & 0.261 \\
      \hline
    \end{tabular} \label{tab:cit}
\end{table}
 
\subsection{Preliminaries} \label{sec:assum}

For the purpose of experimentation, we consider a brownout compliant transactional web application (as shown in Fig.~\ref{fig:arch}) that is deployed on Amazon EC2 and use the Wiki07 request trace to simulate the application workload over a 30-day period. The original Wiki07 data is pre-processed and used to construct 30 daily time series with a lag of 1 second by aggregating the number of requests per second (Rps). These individual time series are then merged into a single continuous time series $TS_{1}$ with a lag of 1 second. From this, a second time series $TS_{2}$, with a 1 minute lag, is created by binning the data using the \textit{minute-of-hour} criterion and computing the maximum Rps per minute. The resource requirements at the logic tier are fulfilled by the candidate instance types shown in~Table~\ref{tab:cit}. For each candidate instance type, the sustainable workload capacity for handling requests fully $\alpha$ is estimated using Amazon EC2's instance configuration guide\footnote{\url{https://docs.aws.amazon.com/AWSEC2/latest/UserGuide/ebs-optimized.html}, accessed on 5 March 2020}. 

\subsection{Proactive Infrastructure Control} \label{sec:pic-sim}

Algorithm~\ref{alg:pic} is used for proactive resource allocation in the PIC loop and the experiments are conducted in R using the \textit{forecast} package\footnote{https://cran.r-project.org/web/packages/forecast/forecast.pdf}. The one-step-ahead workload is estimated first, and then used to optimise the corresponding infrastructure resources.

\subsubsection{One-step-ahead Workload Estimation}

A sliding-window approach is used with $TS_2$ for the one-step-ahead workload estimation. In each step, the forecasting model is trained using historical data with a pre-defined look-back horizon, and then validated using the data corresponding to the next hour. The data window is moved forward by a predetermined sliding length and the process is repeated again. The specific settings used in the PIC loop are as follows:
\begin{itemize}
    \item Forecasting techniques $k$ = \textit{Auto-ARIMA, STL, TBATS}
    \item Look-back horizon $l$ = 24 hours (1440 minutes)
    \item Forecast horizon $f$ = 1 hour (60 minutes)
    \item Sliding window size $w$ = 1 hour (60 minutes)
    \item Workload estimation function $\phi$ = \textit{max}
    \item Performance Metric = RMSE
    \item Forecast Values = mean 
\end{itemize}

\subsubsection{Proactive Resource Scaling} 

The DCO strategy is used for resource scaling based on the estimated workload as well as the actual workload for each hour of the evaluation period. The DCO strategy has been implemented using the Minizinc open-source constraint modeling language\footnote{\url{https://www.minizinc.org/}}. 

\begin{figure}[!t]
\centering
\includegraphics[width=1.0\hsize]{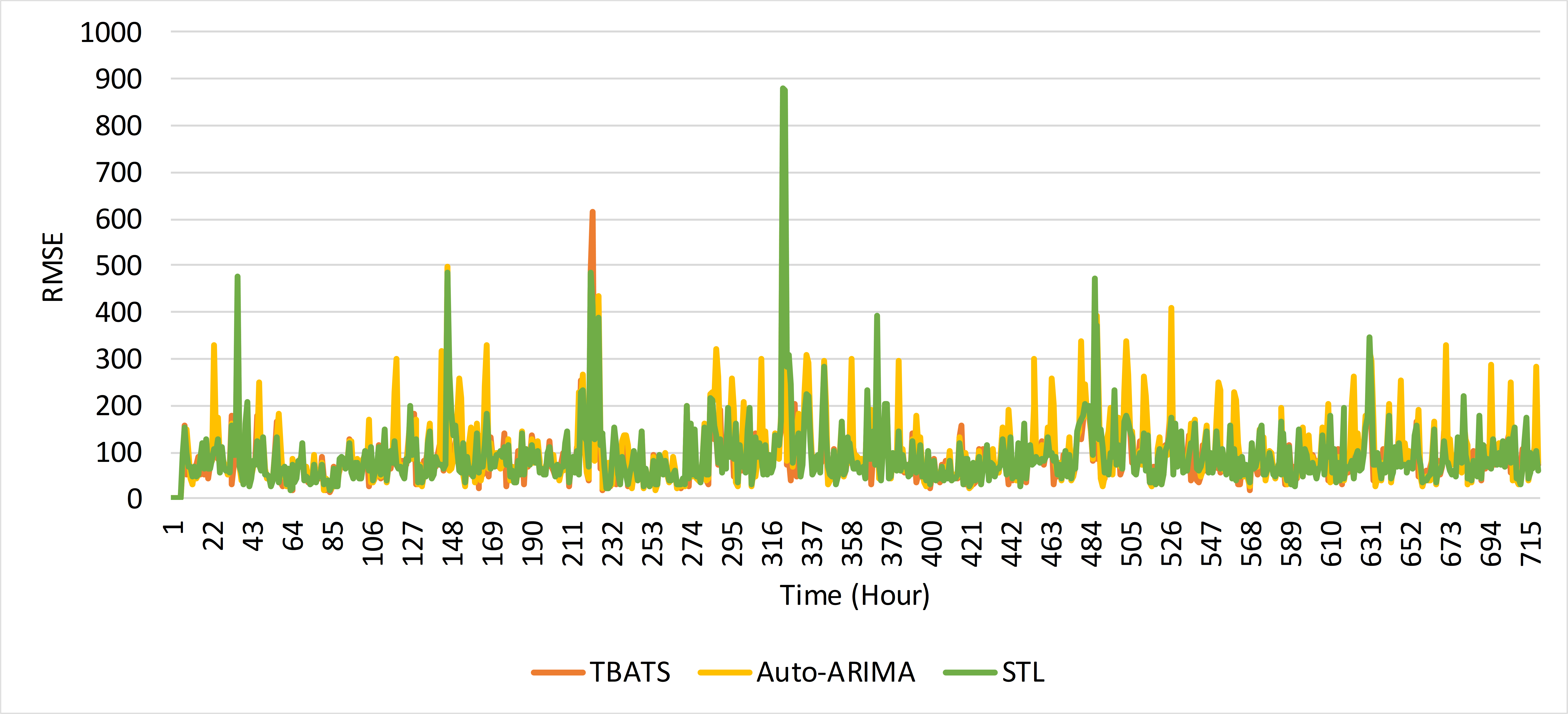}
\caption{RMSE of forecasting techniques over time (PIC Loop).} \label{fig:forper}
\end{figure} 

\begin{table}[!t]
\begin{center}
\caption{Comparison of forecasting algorithms (PIC loop).}
\label{tab:facomp}
\begin{tabular}{c|c|c|c|c|c|c}
\hline
& \textbf{RMSE} & \textbf{BWE} & \textbf{WUE} & \textbf{RUA} & \textbf{BRA} & \textbf{TAC} \\ \hline
TBATS & \textbf{89.67} & 215  & 108 & 42 & \textbf{274} & \$602.45 \\
STL & 91.51 &  204 & \textbf{94} & \textbf{37} & 136 & \$602.47 \\
Auto.ARIMA & 102.95  &  \textbf{301} & 118 & 71 & 236 & \textbf{\$593.18} \\
\hline
\end{tabular}
\end{center}
\end{table}   

\subsubsection{Observations}
\begin{figure}[!t]
\centering
\includegraphics[width=1.0\hsize]{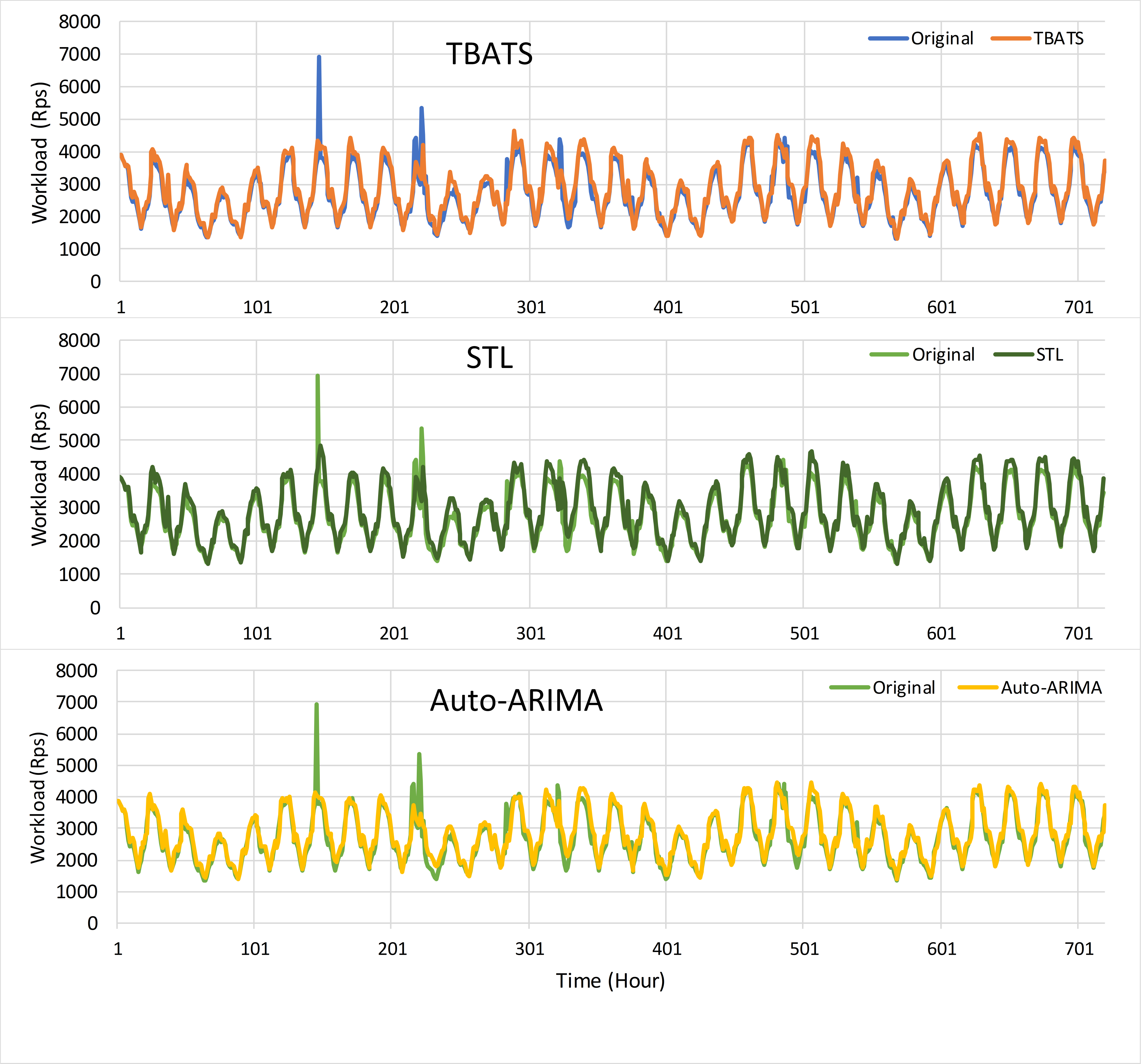}
\caption{Actual Vs. Estimated Workload (PIC loop).} \label{fig:forecasts}
\end{figure}

PIC was simulated for all 720 hours of the evaluation period. As shown in Table~\ref{tab:facomp}, there is no clear winner among the three forecasting techniques in terms of workload estimation. Each technique performs slightly better than the others depending upon which evaluation metric considered. TBATS has the lowest mean RMSE over the 720 hour evaluation period (see~Fig.~\ref{fig:forper}); Auto-ARIMA has the highest number of \textit{best workload estimations} (BWE)\footnote{BWE is measured as the lowest absolute difference between estimated and real maximum workload per hour over the 720 hour evaluation period.}; and STL has the lowest number of \textit{workload under-estimations} (WUE). All three techniques are unable to predict some workload spikes (e.g., \#146 and \#221) (see~Fig~\ref{fig:forecasts}).

Similarly, there is no clear winner in terms of resource allocation resulting from the estimated workloads. STL has the lowest number of resource under-allocations (RUA); TBATS ranks first in terms of the best resource allocation (BRA)\footnote{BRA is measured as the absolute difference between actual workload and aggregate sustainable capacity of allocated resources for each hour of the evaluation period.}; and Auto-ARIMA has the lowest total allocation cost (TAC). All three techniques return the exact same allocation in 74 of the 720 hours.  

\begin{table}[!t]
  \begin{center}
    \caption{Resource Allocation using different $\phi$ (PIC Loop).}
    \label{tab:tac}
    \begin{tabular}{c|c|c|c|c}
    \hline
    \textbf{$\phi$} & \textbf{TAC (\$)} & \textbf{UA} & \textbf{UA} & \textbf{OA} \\
    & & & ($<$ 200 Rps) & ($>$ 200 Rps)\\
    \hline
    \textit{Actual} &  563.42 & & & \\ 
    \hline
    \textit{max} & 602.45 & 42 & 16 & 428\\ 
    90\textsuperscript{th}\%tile & 592.40 & 55 & 19 & 387\\ 
    80\textsuperscript{th}\%tile & 589.50 & 81 & 17 & 323\\ 
    70\textsuperscript{th}\%tile & 584.73 & 107 & 18 & 256\\ 
    60\textsuperscript{th}\%tile & 583.33 & 135 & 22 & 220\\ 
    50\textsuperscript{th}\%tile & 570.13 & 162 & 31 & 189 \\ 
    \hline
    \end{tabular}
  \end{center}
\end{table}   

\subsubsection{Further Experiments}

\begin{figure}[!t]
\centering
\subfloat[Workload Estimation]
{\includegraphics[width=1.0\hsize]{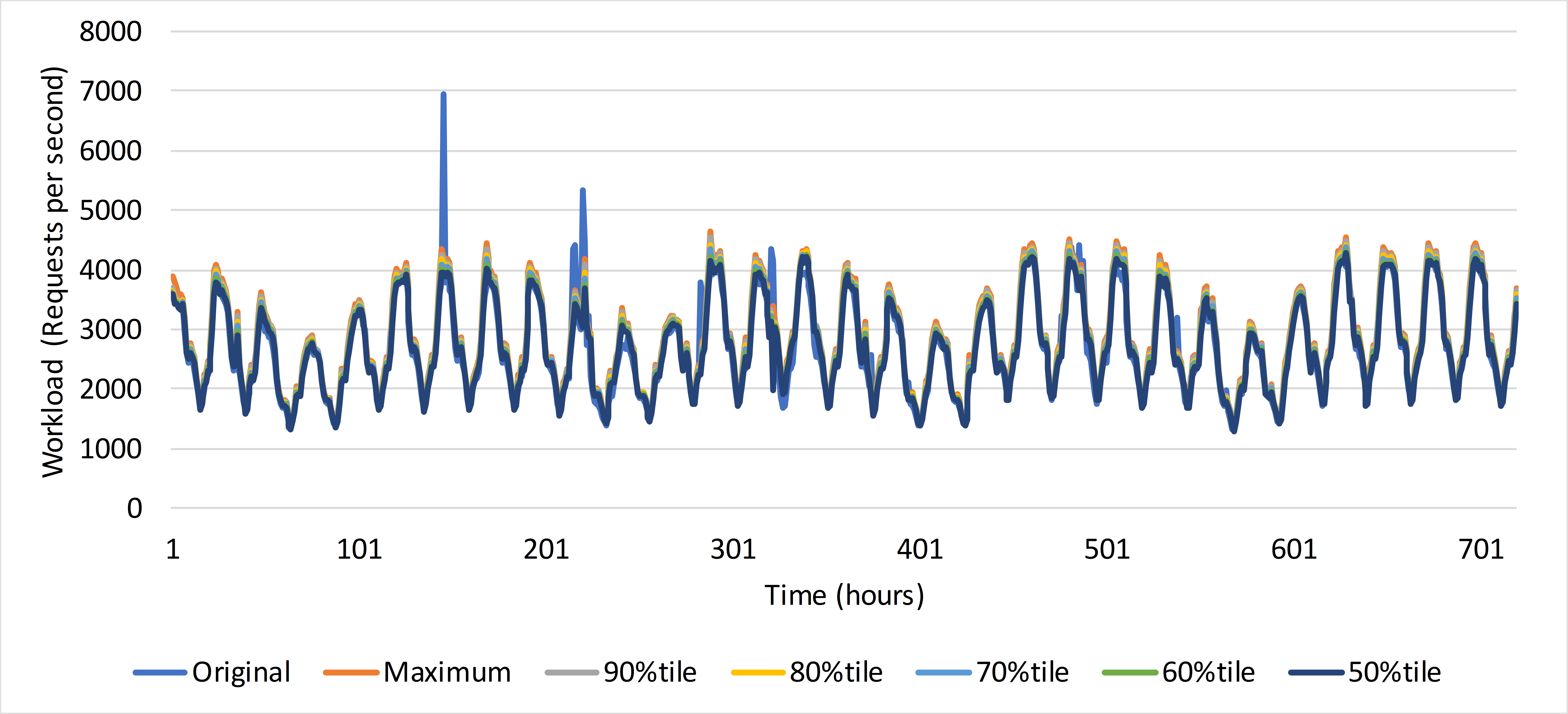}\label{fig:tbats}} \\
\subfloat[Resource Allocation]{\includegraphics[width=1.0\hsize]{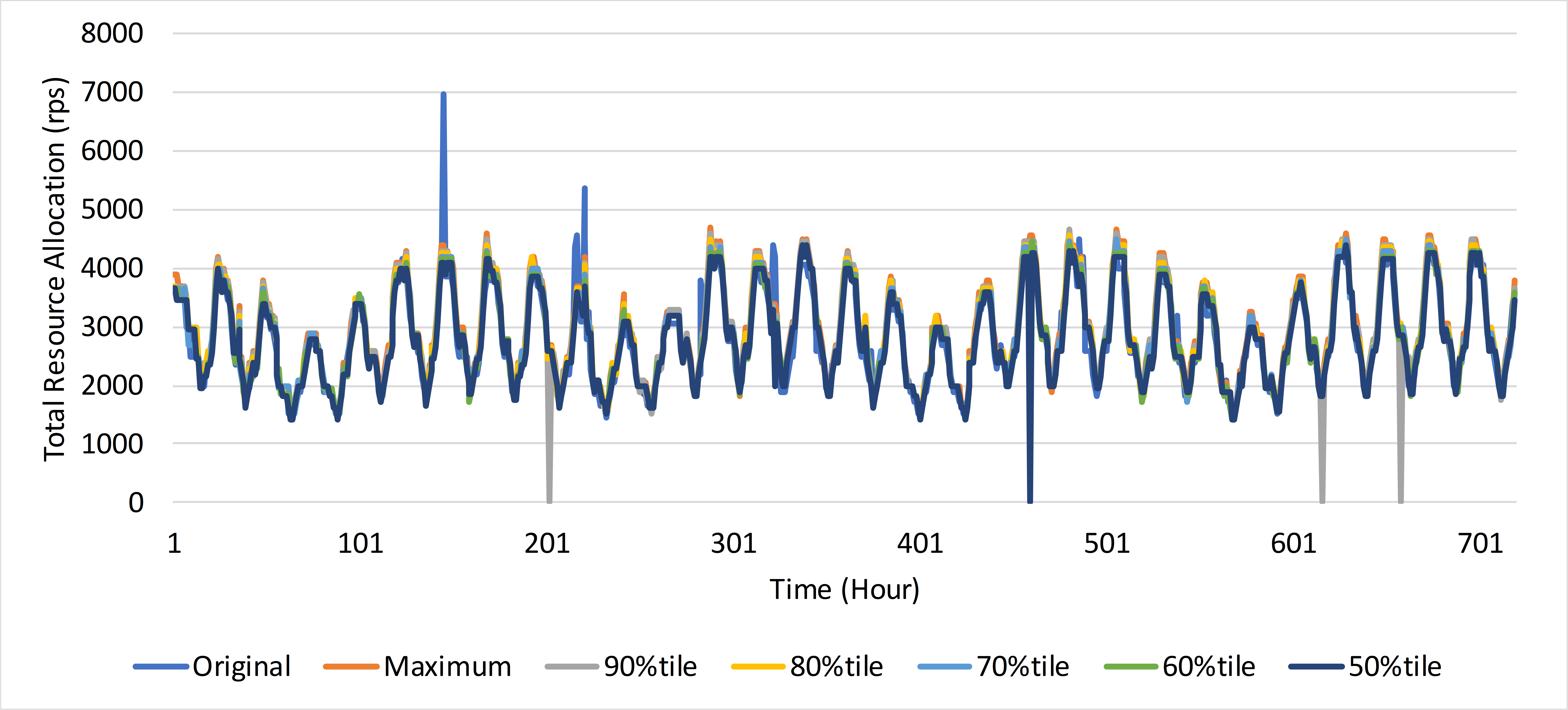}\label{fig:tbats_alloc}}
\caption{TBATs algorithm with different $\phi$ (PIC Loop).}
\label{fig:diffPhi}
\end{figure}

Further experiments were conducted using TBATS with different values for $\phi$ to study the impact on the allocation efficiency and TAC; Table~\ref{tab:tac} shows the result summary; Fig.~\ref{fig:tbats} and Fig.~\ref{fig:tbats_alloc} show the workload estimation and corresponding resource allocation respectively. 
As expected, the number of under allocations increase as $\phi$ moves from \textit{max} to a smaller percentile-based allocation. $\phi$ = \textit{max} returns 42 hours of under allocation compared to 162 hours when $\phi$ = \textit{50\textsuperscript{th}\%tile}.
Similarly, the number of over allocations, where the workload exceeds the aggregate sustainable capacity of the underlying resources, decrease as $\phi$ moves from \textit{max} to a smaller percentile-based resource allocation. With $\phi$ = \textit{50\textsuperscript{th}\%tile}, resource over allocation occurs in 189 hours compared to 428 hours when $\phi$ =  \textit{max}, but the cost savings obtained are rather modest, e.g., allocating resources at the 50\textsuperscript{th}\%tile of the predicted workload returns a modest saving of 5.365\% only. However, it should be noted that only On-Demand instances were used for the simulations and the cost-savings would be much more significant if Spot instances are also considered as demonstrated in \cite{chhetri2019exploiting}. 


    

\subsection{Responsive Software Control} \label{sec:rsc-sim}

The first step in the RSC loop is to determine the one-step-ahead workload $\hat{\omega}_{t+}$, and the second step is to determine the $\zeta_{t+} \mid \eta_{t+} \mid \theta_{t+}$ split. In the PIC loop, resource allocation using TBATS and $\phi$ = \textit{max} returns 42 hours in which the aggregate sustainable capacity of the allocated resources is less than the corresponding real maximum workload. We simulated RSC for those 42 hours. 

\subsubsection{One-step ahead workload estimation}

$TS_1$ is used for workload prediction in the RSC loop with the following settings:
\begin{itemize}
    \item Forecasting technique $k$ = TBATS
    \item Look-back horizon $l$ = 60, 30, 15, 10, 5 minutes
    \item Forecasting horizon $f$ = 60, 30, 15, 10, 5, 1 second
    \item Sliding window size $w$ = 60, 30, 15, 1 second
    \item Forecast Values = mean 
\end{itemize}

Through manual tuning, the most accurate workload prediction was obtained with \textit{$l$ = 1800 sec, $f$ = 1 sec, $w$ = 1 sec.}. 

\subsubsection{Responsive Software Scaling}

Algorithm~\ref{alg:rsc} is applied to the predicted workload to determine the optimal $\zeta_{t+} \mid \eta_{t+} \mid \theta_{t+}$ split. Table~\ref{tab:rsc} provides a summary of the RSC simulation results. The total number of results above the sustainable workload capacity $\alpha_T$ in the 42 hour period is \textbf{190581}, i.e., \textit{without RSC, 190581 requests cannot be handled normally} and this has an adverse impact on the application performance. Using RSC, the estimated $\zeta_{t+} \mid \eta_{t+} \mid \theta_{t+}$ split is \textit{340888135}|\textit{798015}|\textit{1805} based on the predicted workload (\textit{345886150} in total). However, since the workload prediction is not 100\% accurate, the ``corrected'' workload split is \textit{334724348}| \textit{527134}|\textit{17836}. By combining PIC with RSC, \textit{the system is able to accommodate \textbf{90.64\%} of the requests that could not be previously handled} using PIC only.  
 
\begin{figure}[!t]
	\centering
	\subfloat[Real vs Forecast Workload]
	{\includegraphics[width=1.0\hsize]{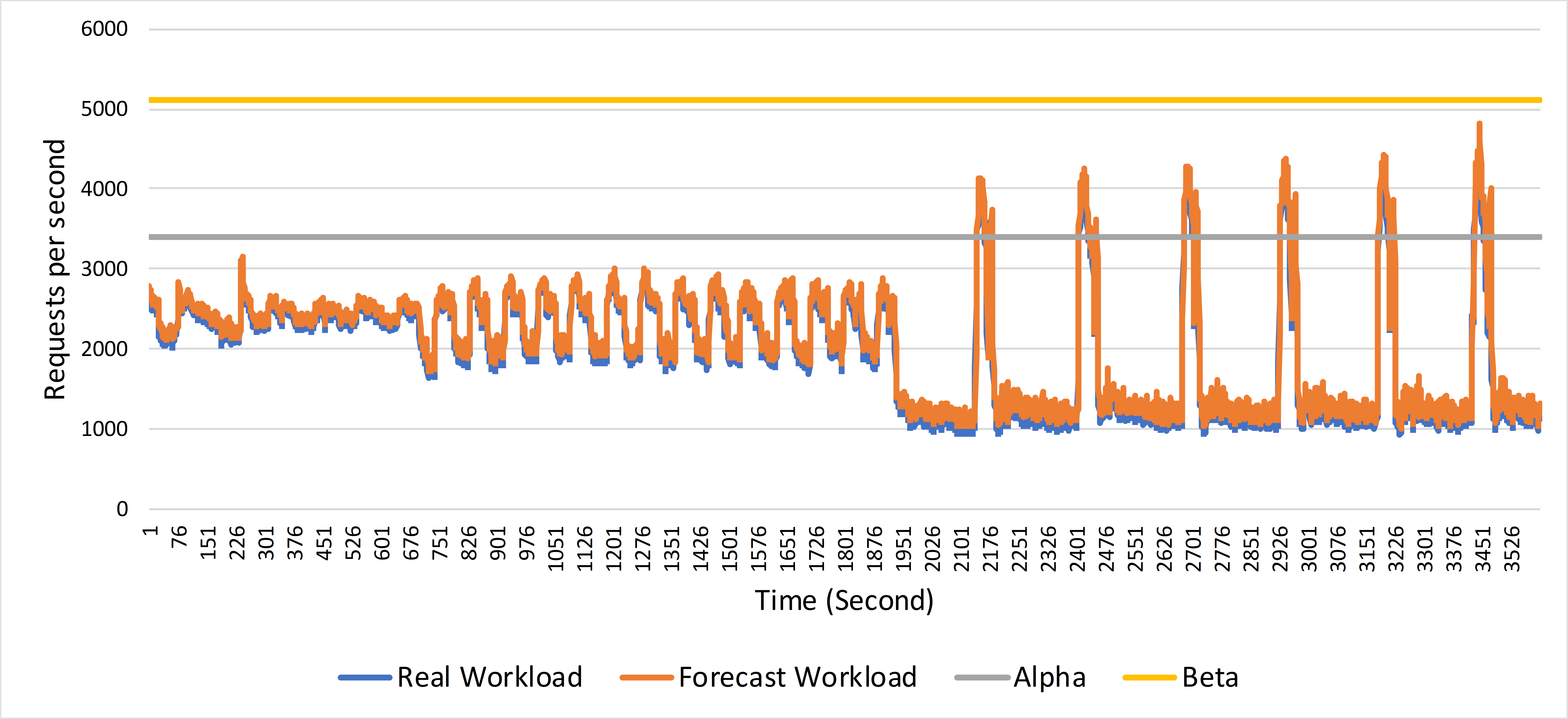}\label{fig:rsc_a1}} \\
	\subfloat[Total sustainable capacity using RSC ($\gamma$ + $\varphi$)]
	{\includegraphics[width=1.0\hsize]{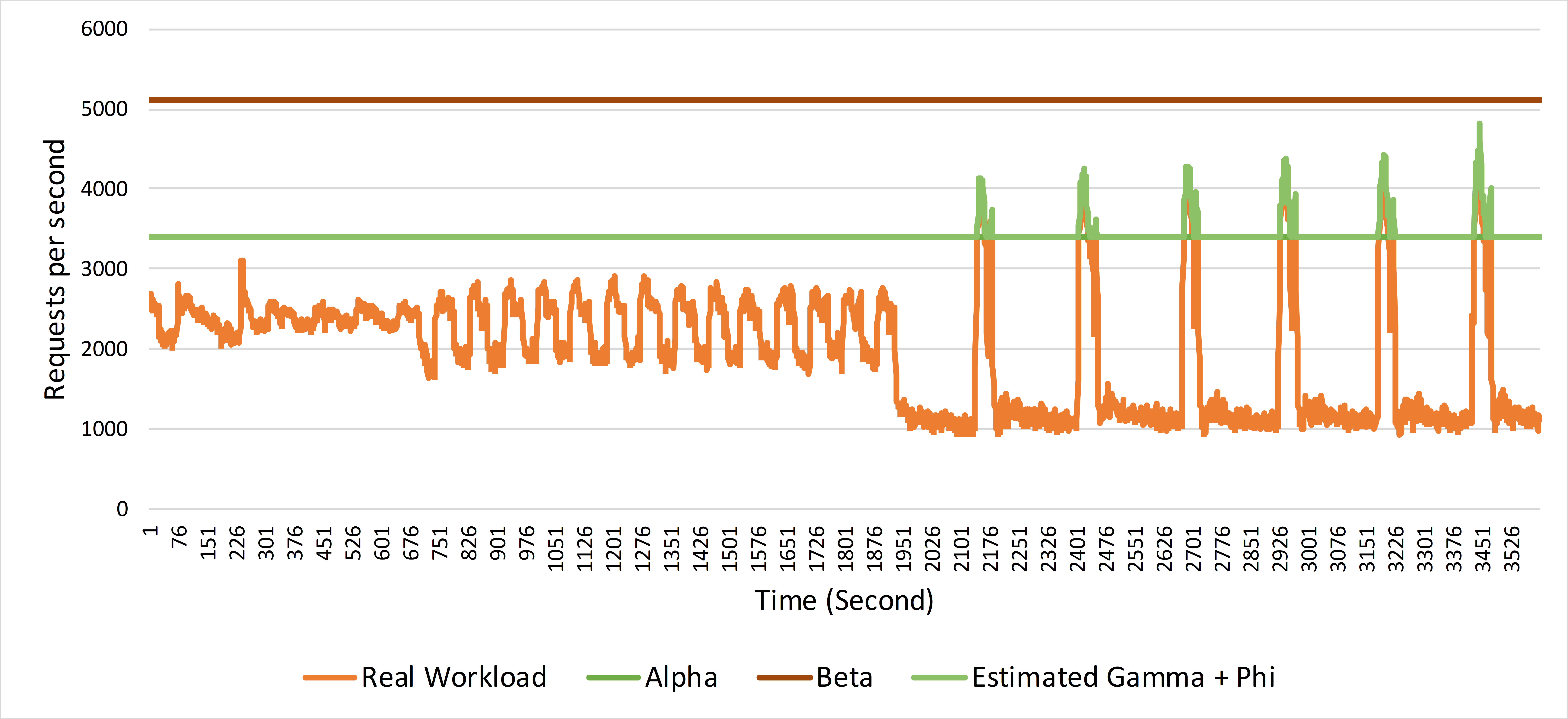}\label{fig:rsc_b1}}
	\caption{RSC in Scenario 2 (\#322).}
	\label{fig:rsc2}
\end{figure}

To illustrate the impact of prediction accuracy on the effectiveness of RSC, we refer to hour \#322, which has the highest number of instances in which the actual workload $\omega_{t}$ exceeds $\alpha_{322}$ over the 30-day evaluation period. The aggregate sustainable capacity $\alpha_{322}$ = 3400 Rps and the workload increases are concentrated in six regions (cf.~Fig.~\ref{fig:rsc_a1}). Of the 7067197 requests received, a total of 66402 requests cannot be processed due to infrastructure resource limitations. We present the results under two specific settings: 
 
\begin{table}[!t]
\begin{center}
\caption{Result Summary (RSC Loop).}
\label{tab:rsc}
\begin{tabular}{l | r}
\hline
Total requests received in the 42 hours & 335269318 \\
\hline 
Number of requests above $\alpha_{T}$ & \textbf{190581} \\
\hline
\textbf{Estimated Split (based on $\hat{\omega}_{t+}$}) & \\
$\zeta_{t+}$ & 340888135\\
$\eta_{t+}$ & 798015 \\
$\theta_{t+}$ & 1805 \\
\hline
\textbf{Corrected Split (based on $\omega_{t+}$}) & \\
$\zeta_{t+}$ & 334724348 \\
$\eta_{t+}$ & 527134\\
$\theta_{t+}$ & \textbf{17836} \\
\hline
\end{tabular}
\end{center}
\end{table}  
 
\begin{itemize}[leftmargin=*]
    \item \textbf{Scenario 1} 
    (\textit{$l$ = 3600 sec, $f$ = 60 sec, $w$ = 60 sec.)}\textbf{:} Fig.~\ref{fig:rsc_a} shows the predicted workload $\hat{\omega}_{t+}$ against the actual workload $\omega_{t+}$. TBATS completely misses three regions (2, 4 and 5) in which the workload $\omega_{t+}$ exceeds $\alpha_{322}$. As per the prediction, there are 66 seconds in total in which $\hat{\omega}_{t+} > \alpha_{322}$. Following the redistribution of $\hat{\omega}_{t+}$ between $\zeta_{t+}$ and $\eta_{t+}$, 20211 additional requests can be handled (30.44\% of previously unhandled requests). However, RSC does not handle all 66402 requests because of the poor workload prediction. It should be noted that the workload redistribution is based on the predicted workload and not the actual workload. 
    
\begin{figure}[!t]
\centering
\subfloat[Real vs Forecast Workload]
{\includegraphics[width=1.0\hsize]{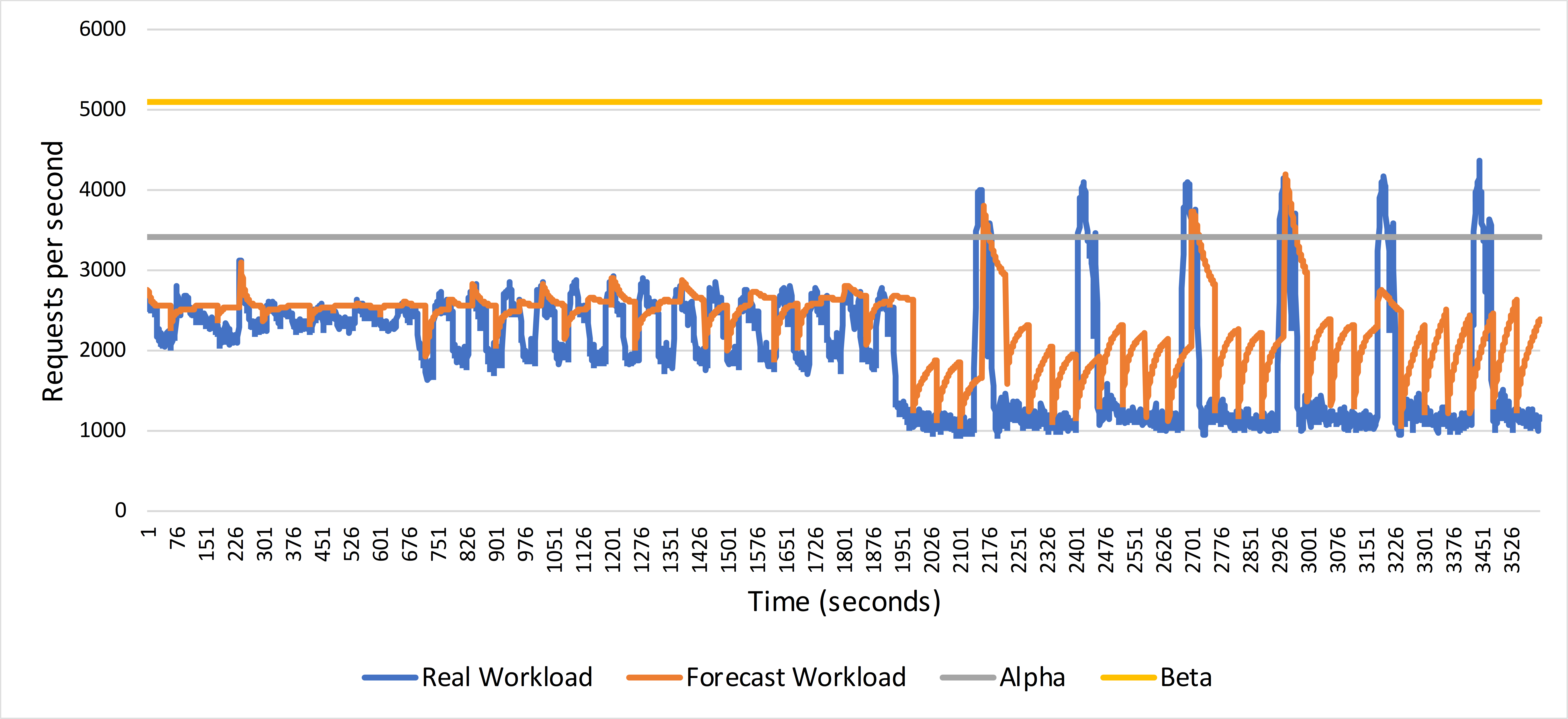}\label{fig:rsc_a}} \\
\subfloat[Total sustainable capacity using RSC ($\gamma$ + $\varphi$)]{\includegraphics[width=1.0\hsize]{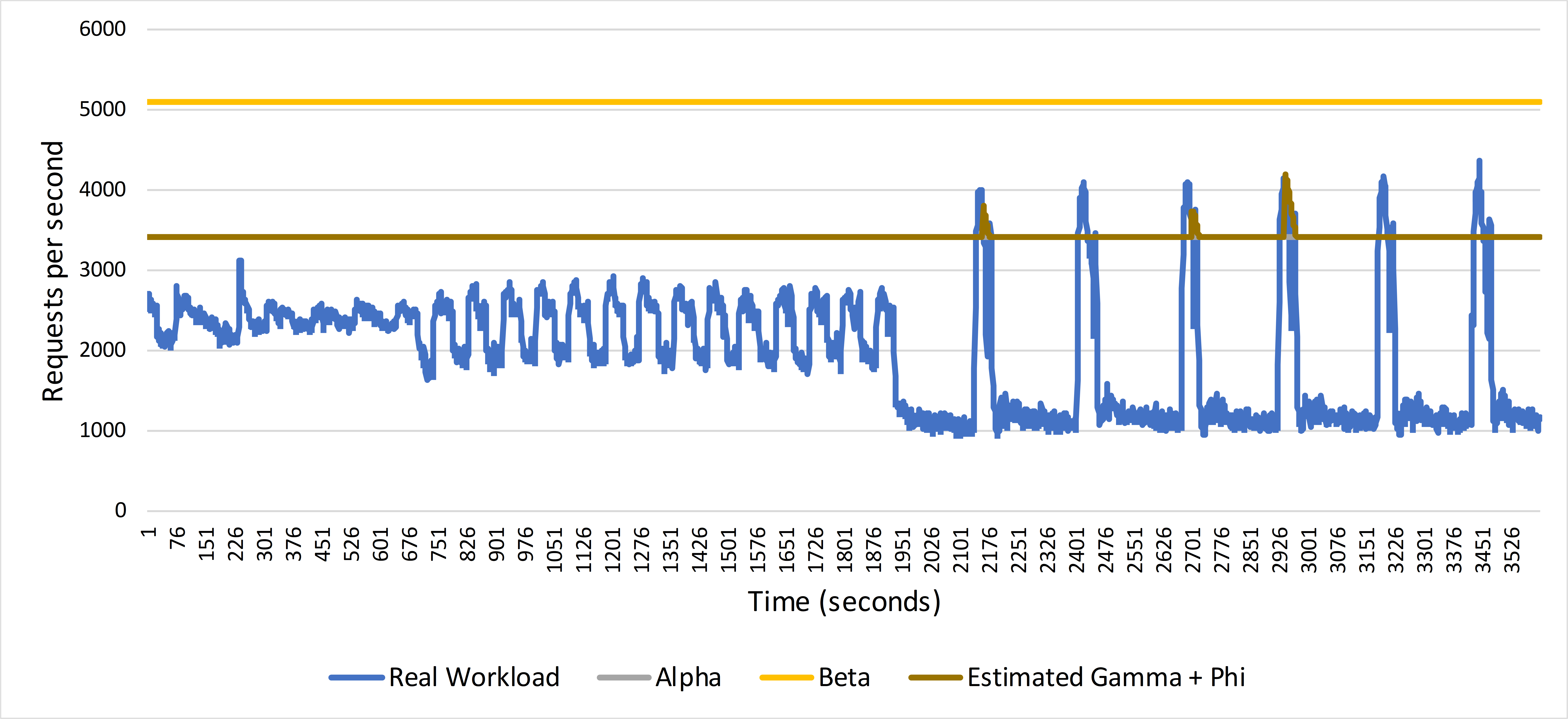}\label{fig:rsc_b}}
\caption{RSC in Scenario 1 (\#322).}
\label{fig:rsc1}
\end{figure}

    \item \textbf{Scenario 2} (\textit{$l$ = 1800 sec, $f$ = 1 sec, $w$ = 1 sec.})\textbf{:} Fig.~\ref{fig:rsc_a1} shows the predicted workload $\hat{\omega}_{t+}$ against the actual workload $\omega_{t+}$. With the selected settings, TBATS is able to accurately pick all 6 regions where the workload exceeds $\alpha_{322}$. More specifically, the workload is predicted to exceed $\alpha_{322}$ 208 times. Following the redistribution of $\omega_{t+}$ among $\zeta_{t+} \mid \eta_{t+} \mid \theta_{t+}$, the resource capacity allocated by the PIC loop can handle an extra 66301 requests, \textit{i.e.,} 99\% of the previously unhandled requests, in the RSC loop.  
\end{itemize}


The results highlight that combining RSC with PIC always improves the application performance, although, similar to PIC, the effectiveness of RSC also depends upon the prediction accuracy. 


\subsection{Reactive Infrastructure Control} \label{sec:ric-sim}

As explained in Section~\ref{sec:ris} (RIC loop realization), the scaling policy determines \textit{when} reactive scaling is triggered, but not \textit{how much}. The actual resource allocation is determined by the observed delta change in the workload. RIC for scale-out\footnote{Evaluation of reactive scale-in has been omitted due to space limitations.} was simulated over the 720-hour evaluation period using the following rule -- \textit{trigger reactive scale-out if the workload exceeds the threshold of 200 Rps\footnote{In our simulations, the minimum sustainable capacity among all instances in the candidate pool is 200 Rps for \textit{t2.medium}.} at least once in 3 consecutive evaluation periods of 5 minutes duration}. 

\begin{figure}[!t]
\centering
\includegraphics[width=1.0\hsize]{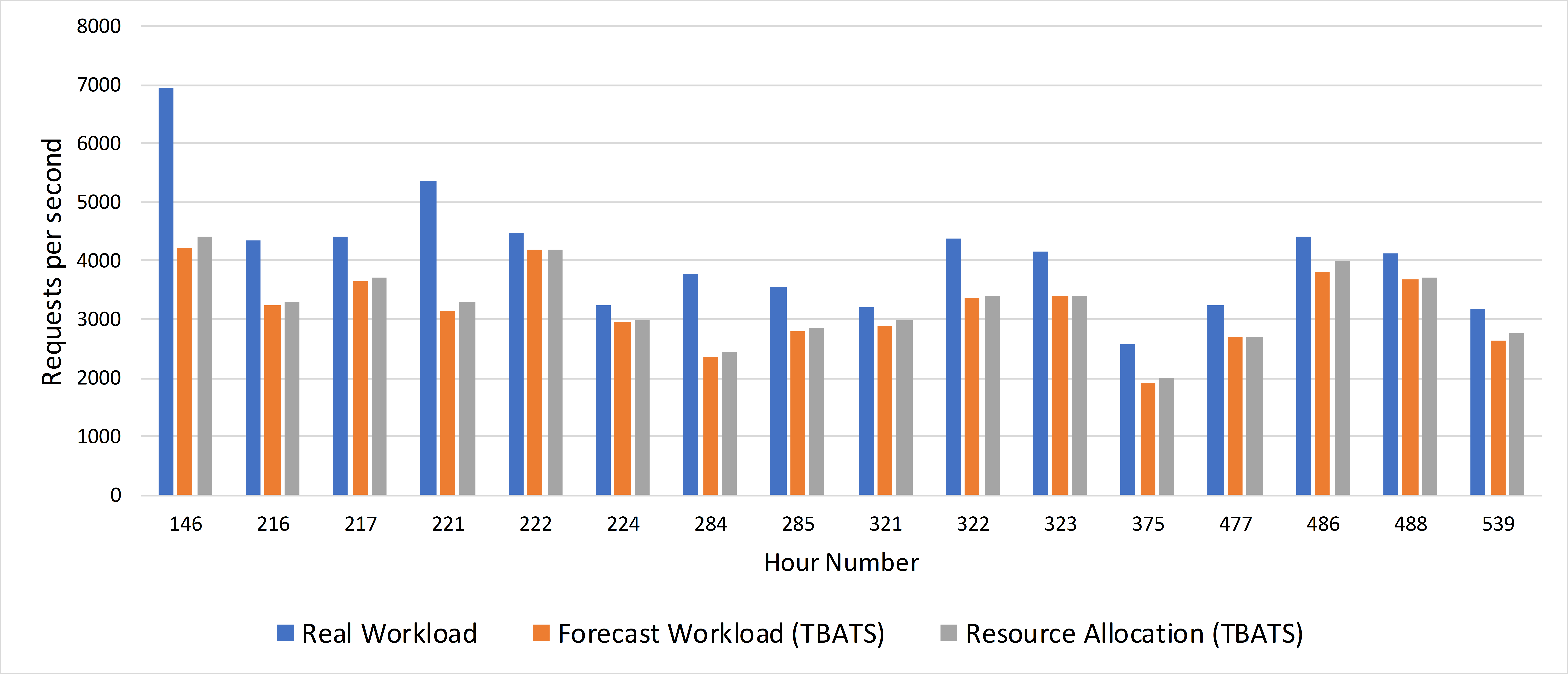}
\caption{Real Workload, Predicted Workload, $\alpha$ (RIC Loop).} \label{fig:ric}
\end{figure}

There are 16 hours (out of 720) in which the actual workload $\omega_t$ exceeds the corresponding sustainable capacity $\alpha_T$ by 200 Rps or more (cf.~Fig.~\ref{fig:ric}). However, reactive scaling is triggered only in 3 hours, i.e., \#221, \#322, and \#323. 
For example, in \#221, the average workload above $\alpha_{221}$ over the 3 continuous observation periods is 673 Rps, and the maximum workload is 2049 Rps. Reactive resource allocation based on the average workload increase (Avg $\omega_{t+}$) results in 3 instances (2 \textit{t2.medium} and 1 \textit{c4.large}) being added to the existing capacity at a cost of \$0.188. Similarly, resource allocation based on the maximum workload increase (Max $\omega_{t+}$) results in 6 instances (3 \textit{t2.medium} and 3 \textit{c4.large}) being added at a cost of \$0.565. Table~\ref{tab:ric} summarises the results for all three hours. 

Decisions taken in the RSC have no impact on the decisions made in the PIC and RIC loops. However, decisions made in the RIC loop have an immediate impact on RSC decisions in the current time step $T$ of the PIC loop as well as PIC decisions in the subsequent time step $T+$. 

\begin{table}[!t]
  \centering
   \caption{Result Summary (RIC Loop)} 
    \begin{tabular}{c|c|c|c|c} 
    \hline
      \textbf{T} & \textbf{$\mathcal{R}_T$} & \textbf{Avg/Max $\omega_{t+}$} & $\mathcal{R}_{T+ |\textbf{Avg/Max} \omega_{t+}}$ & \textbf{AC (\$)}\\
      \hline
      \#221 & 3300 & 673/2049 & 700/2100 & 0.188/0.565\\
      \#322 & 3400 & 383.8/964 & 400/1050 & 0.117/0.300\\
      \#323 & 3400 & 374/755 & 400/900 & 0.117/0.025\\
      \hline
    \end{tabular} \label{tab:ric}
\end{table}

\subsection{Discussion}

Even though our experiments evaluated the three auto-scaling loops separately, they confirm that the combined use of PIC, RSC, and RIC can achieve better outcomes in terms of request handling performance than the isolated use of each loop. They also confirm that there is ample scope for realizing self-improvement in all three control loops. In particular, our PIC and RSC experiments indicate that varying the parameters of a specific prediction algorithm or using different algorithms over time would be beneficial; and this could be achieved by executing them in parallel and learning which one supports increased workload performance under different conditions. Indeed, the coordinator loop could use reinforcement learning (RL) in order to explore and exploit the parameter space (time horizons, etc.) for optimum throughput and responsiveness. 

Our additional PIC experiments suggest there is also value in automatically tuning the auto-scaling responsibility or ``optimal split'' between the PIC and RSC loops, e.g., the PIC could underallocate resources to minimize costs and leave the RSC loop to handle the remaining deviations -- something that can also be effected by a controller loop as part of self-improvement. We have previously demonstrated that significant cost savings can be achieved by jointly leveraging dedicated (e.g., On-Demand instances) and transient (e.g., Spot instances) resources~\cite{chhetri2019exploiting}.


\section{Conclusion} \label{sec:conc}

In this paper, we presented a novel approach for elasticity in cloud-based software systems that leverages and builds on infrastructure and software auto-scaling and proactive, reactive and responsive decision-making. We also presented a conceptual architecture for a self-improving elasticity controller based on our approach and the popular MAPE-K reference model, and a partial realization of this architecture inclusive of algorithms for proactive, responsive and reactive auto-scaling. We evaluated our controller realization via simulation with real-world workload datasets. Experimental results, although conducted on each control loop in isolation, demonstrate the benefits of combining infrastructure and software elasticity control through proactive, reactive and response decision-making. Our future work will consider full realization of the coordinator and meta-coordinator loops, as well as experimentation using both dedicated and transient cloud resources. We will also investigate how these ideas could be applied for resilient cloud resource management, both in the presence of load dynamics as well as cloud infrastructure revocation or failure. 

\bibliographystyle{ACM-Reference-Format}
\bibliography{bibfile}

\end{document}